\documentclass{iopconfser}

\usepackage{caption}
\usepackage{subcaption}

\usepackage[numbers]{natbib}
\usepackage{url}
\usepackage{lmodern}
\usepackage{amsmath}
\usepackage{amssymb}
\usepackage{mathtools}
\usepackage{cleveref}

\usepackage{tikz}

\def\equationautorefname~#1\null{Equation~(#1)\null}
\newcommand{\isotope}[2]{\(^{#2}\mathrm{#1}\)}
\newcommand{\unit}[1]{\, \mathrm{#1}}

\begin{document}

\title{Approximate Differentiable Likelihoods for Astroparticle Physics Experiments}

\author{Juehang Qin$^{1}$ and Christopher Tunnell$^{1}$}

\affil{$^1$Department of Physics and Astronomy, Rice University, Houston, TX, USA}

\email{qinjuehang@rice.edu}

\begin{abstract}
Traditionally, inference in liquid xenon direct detection dark matter experiments has used estimators of event energy or density estimation of simulated data. Such methods have drawbacks compared to the computation of explicit likelihoods, such as an inability to conduct statistical inference in high-dimensional parameter spaces, or a failure to make use of all available information. In this work, we implement a continuous approximation of an event simulator model within a probabilistic programming framework, allowing for the application of high performance gradient-based inference methods such as the No-U-Turn Sampler. We demonstrate an improvement in inference results, with percent-level decreases in measurement uncertainties. Finally, in the case where some observables can be measured using multiple independent channels, such a method also enables the incorporation of additional information seamlessly, allowing for full use of the available information to be made.
\end{abstract}

\section{Introduction}

Inference in particle physics is often based on black-box simulators that can faithfully reproduce the physics involved; however, such simulations do not provide tractable likelihoods or probability densities, leading to the use of either Approximate Bayesian Computation or approximate likelihoods obtained via density estimation~\cite{Cranmer_2020}. Simulators constructed within probabilistic programming languages
present an alternative that can allow us to conduct simulation-based inference by computing differentiable likelihoods directly using the models that our simulators are based on~\cite{Cranmer_2020}. 

Prior work in particle physics which integrates existing event simulators into the probabilistic programming paradigm, but without making the simulators directly compatible with automatic differentiation~\cite{Baydin2018EfficientPI, Baydin_2019}. Making differentiable models is often challenging in particle physics as many physical processes, such as generation of quanta, are discrete. 
Despite these difficulties, compatibility with automatic differentiation frameworks such as \texttt{JAX}~\cite{jax2018github} confers significant benefits, enabling the application of high-performance gradient-based inference methods, such as variational inference~\cite{Hoffman2012StochasticVI, wingate2013automated}, and the No-U-Turn Sampler (NUTS)~\cite{Hoffman2011TheNS}. 
In this work, we demonstrate an approach using an approximate binomial distribution and the probabilistic programming framework \texttt{numpyro}~\cite{bingham2019pyro, phan2019composable} to compute differentiable likelihoods. To that end, we introduce bayesNEST, a partial implementation of the NEST model for liquid nobel element particle physics detectors, including dark matter experiments XENON and LZ~\cite{XENON:2020kmp, LZ:2019sgr}.

Dark matter is one of the greatest mysteries in physics--while most of the matter in the universe is dark matter, its particle nature is not understood~\cite{ParticleDataGroup:2022pth}. Consequently, liquid xenon detectors are built to search for WIMP dark matter~\cite{Aalbers:2022dzr}. These experiments have other physics channels, including measurements of double electron capture in \isotope{Xe}{124}~\cite{XENON:2019dti, XENON:2022evz}. 
These measurements are typically conducted in a low-statistics regime, necessitating sophisticated approaches to inference. 
Typically, an estimator for event energy is used for inference when searching for electronic recoil signals in these liquid xenon experiments
~\cite{XENON:2020rca, XENON:2022ltv}. In analyses using nuclear recoil events, likelihoods are instead estimated using density estimation from simulated events~\cite{XENON:2019izt}. Density estimation techniques suffer from the curse of dimensionality~\cite{Cranmer_2020}, whereas estimators computed from observed data might not make use of all available information.

Another approach to this problem is to marginalise over all discrete parameters the likelihood~\cite{James:2022sgg}. While this approach is more exact, it can be very computationally and memory intensive to construct the large arrays of likelihoods needed to enumerate over the plausible range of discrete parameters. As such, this work is complementary and can be used to reduce the computational and memory requirements of inference using explicit likelihoods where the approximations we introduce are acceptable.

\section{Background}
\subsection{Dual-phase Xenon Time Projection Chambers}

Dual-phase xenon time projection chambers (TPCs) use an active target of liquid xenon, with a layer of gaseous xenon in the upper portion of the detector. The detector is instrumented with arrays of photomultiplier tubes (PMTs) on the top and bottom. When a particle interaction occurs in the liquid xenon target, scintillation photons are produced. This signal, termed S1, is detected by the PMTs. In addition, electrons are liberated and drifted towards the gaseous region at the top of the detector using an electric field, the drift field. These electrons produce a secondary S2 signal in the gaseous phase.

In the current-generation dual-phase xenon detectors experiments LZ and XENONnT, there are $494$ PMTs distributed between the top and bottom sensor arrays~\cite{LZ:2015kxe, XENON:2020kmp}.
Signals are classified as S1 or S2 based on the shape of the signal waveform and the distribution of the signal across the sensors~\cite{XENON:2022ltv, LZ:2022lsv}, and the amplitude of signals are determined by integrating the signal waveforms. These TPCs are capable of 3D position reconstruction; $(x,y)$ positions can be estimated using the pattern of S2 signal amplitudes across the top PMT array, whereas the $z$ positions can be estimated from the time delay between the S1 and S2 signals. Events can be separated into several categories, including electronic recoils, typically produced by gamma rays and beta decays, and nuclear recoils, produced by neutrons, neutrinos, and WIMP dark matter. More details about liquid xenon TPCs can be found in the following review:~\citep{Aalbers:2022dzr}. 

\subsection{The Signal Generation Model}\label{ssec:NEST}

In this work, we focus on inference involving electronic recoil events, and thus have implemented a probabilistic model for the S1 and S2 signals expected from beta decays; our implementation will be described in~\cref{ssec:bayesNEST_model}. In this section, we will describe how these signals are modelled for electronic recoil events according to the Nobel Element Simulation Technique (NEST). This description is based on~\cite{szydagis_2023_7577399} and codebase of \texttt{NEST} version 2.3.11~\cite{szydagis_2023_7577399}.

Given an interaction of energy $E$, the total number of quanta is computed as the following truncated gaussian discretised by rounding,
\begin{equation}
    \mu_q = \frac{E}{W}, \quad N_{q} \sim \left\lfloor\mathrm{T}\mathcal{N}\left(\mu_q, \sqrt{\mu_q}, a=0, b=\infty\right) + 0.5\right\rfloor,
\end{equation}
where $W [\mathrm{eV}]=21.94 - 2.93\rho$ is the workfunction, the average energy required to produce a single quanta, and $(a, b)$ is the support of the truncated distribution. $\rho$ refers to the density of liquid xenon. Typical values of the workfunction lie within the range of $13\unit{eV} \sim 14\unit{eV}$.

The quanta $(N_q)$ are sorted into the categories of excitons and ions using a binomial distribution, $N_i \sim \mathrm{B}(N_q, p_i)$, $N_{ex} = N_q - N_i$,
where $N_i$ is the number of ions, $N_{ex}$ is the number of excitons, and $p_i$ is the probability of each quanta being an ion. Finally, some ions represented by $N_i$ might undergo recombination to be produce S1 photons, instead of being part of the S2 ionisation signal. This is modelled using a discretised and truncated skew-normal distribution, as given by
\begin{equation}
    \mu_e = (1-p_r)N_i, \quad
    N_e \sim \left\lfloor\mathrm{TSkewN}\left(\mu_e, \sigma_e, \alpha, a=0, b=\infty\right) + 0.5\right\rfloor,\quad
    N_{\gamma} = N_q - N_e,
\end{equation}
where $p_r$ is the recombination probability, $\mu_e$ and $\sigma_e$ are the mean and standard deviation of the skew-normal distribution, and $\alpha$ is the shape parameter controlling skewness.

From $N_{\gamma}$, the number of photons, and $N_e$, the number of electrons, the S1 signal simulated using the following detector model:
\begin{equation}
    N'_{\gamma} \sim \mathrm{B}\left(N_{\gamma}, \frac{g1}{p_{DPE}}\right),\quad
    N''_{\gamma} \sim \mathrm{B}\left(N'_{\gamma}, p_{DPE}\right),\quad
    S1 \sim \mathcal{N}\left(N'_{\gamma} + N''_{\gamma}, \sigma_{PS}\sqrt{N'_{\gamma} + N''_{\gamma}}\right)
\end{equation}
where $\frac{g1}{p_{DPE}}$ is the probability of detecting a photon, $p_{DPE}$ is the probability of a sensor detecting a doubled signal from a single scintillation photon due to double photoelectron emission~\cite{Faham:2015kqa}, and $\sigma_{PS}$ is the the sensor energy resolution. The S2 detector model is instead given by
\begin{equation}
    N'_e \sim \mathrm{B}\left(N_{e}, p_e\right),\quad
    S2 \sim \mathcal{N}\left(N'_e, \sigma_e\sqrt{N'_e}\right),
\end{equation}
where $p_e$ is the probability that an electron is successfully drifted to the top of the liquid xenon and then extracted into the gas phase, and $\sigma_e$ is the spread of the S2 signal. It can be seen that many of the latent parameters are discrete,
posing a problem for direct implementation into a differentiable model.
\section{Methods}
\subsection{Continuous Approximation for the Binomial Distribution}~\label{ssec:beta_norm}


Approximations can be used to work around the aforementioned discrete random variables. The normal distribution is a commonly used approximation, but it does not perform well when $p$ is far away from 0.5 as there is no skewness. This can be remedied using either transformations of approximating distributions such as a normal distribution~\cite{Gebhardt:127bc0fe-1b1f-35af-bccd-7687f9f56c9e}, or a skew-normal distribution~\cite{chang8bd310a2-bf97-36f1-b784-d8427f10b7b9}. Finding the skew-normal shape parameters that approximate a binomial distribution, however, requires one to solve an equation where the solution does not have a simple functional form. This makes it tricky to use the skew-normal distribution as an automatic-differentiation-compatible approximation of the binomial distribution. To address this issue, we use a mixture of a normal distribution and a beta distribution scaled by $N$, the number of trials.
We believe that this is the first time a beta-normal mixture distribution has been used as a continuous approximation of a binomial distribution to obtain a differentiable likelihood.

We match the first two moments of the distributions to determine the correct parameters of the normal and beta distributions that approximate a given binomial distribution. If a mixture distribution comprises distributions with equal mean, then the higher-order central moment of the mixture distribution is simply given by a weighted sum of the moments of the constituent distributions. This can be derived as such:
\begin{equation}
    \mathbb{E}\left[(X-\mu)^n\right] = \int_{-\infty}^{\infty}(x-\mu)^n \sum_{i=0}^k w_i p_i(x)
    = \sum_{i=0}^k w_i \int_{-\infty}^{\infty}(x-\mu)^n p_i(x)
    = \sum_{i=0}^k w_i \mathbb{E}\left[(X_i-\mu)^n\right],
\end{equation}
where $X$ is a random variable with the probability density $f(x) = \sum_{i=0}^k w_i p_i(x)$. Thus, if the normal and beta distributions match the mean and variance of the binomial distribution, any mixture of the two will match the first two moments of the binomial distribution. Similarly, the non-central moments of a mixture distribution can also be given by a weighted-sum of the constitutents.


The mean and variance of binomial distribution $\mathrm{B}(n, p)$ are $np$ and $np(1-p)$, respectively~\cite{Walck:1996cca}. This corresponds to the normal distribution $\mathcal{N}[\mu=np, \sigma=\sqrt{np(1-p)}]$. As the beta distribution is the conjugate distribution of the binomial distribution~\cite{hoffalma991033186282805251}, one might intuitively expect the corresponding Beta distribution to be $\mathrm{Beta}(\alpha, \beta)$, where $\alpha=np$, and $\beta=n(1-p)$. However, it turns out that this distribution does not match the variance of the binomial distribution. Instead, the matching beta distributions has the following parameters:
\begin{equation}
\label{eq:corrected_beta_params}
    \alpha' = \alpha - \frac{\alpha}{\alpha + \beta},\quad
    \beta = \beta - \frac{\beta}{\alpha + \beta}.
\end{equation}

The next step is to match the third standardised moment of the distributions, given by
\begin{equation}
    \gamma = \mathbb{E}\left[\left(\frac{X - \mu}{\sigma}\right)^3\right]
    = \frac{\mathbb{E}\left[X^3\right] - 3\mu\sigma^2 -\mu^3}{\sigma^3}.
\end{equation}

It can be seen that the skewness of a distribution does not depend on the scale of a distribution, and thus the following derivation can be carried out without needing to additionally account for the scale of the beta distribution. As $\mathbb{E}\left[X^3\right]$ can also be obtained via a weighted sum for a mixture distribution, and the means and variances are already matched, we can simply focus on this raw moment. The skewness of a normal distribution is zero; as such, $\mathbb{E}\left[X_{\mathcal{N}}^3\right] = 3\mu\sigma^2 + \mu^3$. 
The skewness of $\mathrm{Beta}(\alpha', \beta')$ is $\gamma_\beta = [2(\beta' -\alpha' ){\sqrt {\alpha' +\beta' +1}}]/[(\alpha' +\beta' +2){\sqrt {\alpha' \beta' }}]$~\cite{Walck:1996cca}. The skewness of a binomial distribution is given by $\gamma_\mathrm{Binom} = \frac{1-2p}{\sqrt{Np(1-p)}}$~\cite{Walck:1996cca}.

We can thus find the relative weights by solving the following system of equations for the weight $w$, where $w$ is the weight for the beta distribution in the mixture, and correspondingly the weight for the normal distribution is $1-w$:
\begin{equation}
    \frac{x - 3\mu\sigma^2 -\mu^3}{\sigma^3} = \frac {2(\beta' -\alpha' ){\sqrt {\alpha' +\beta' +1}}}{(\alpha' +\beta' +2){\sqrt {\alpha' \beta' }}},
    \frac{1-2p}{\sqrt{Np(1-p)}} = \frac{w x + (1-w)(3\mu\sigma^2 + \mu^3) - 3\mu\sigma^2 -\mu^3}{\sigma^3}.
\end{equation}
The solution is given by $w = \frac{\alpha' + \beta' + 2}{2 \left(\alpha' + \beta' + 1\right)}$.
\begin{figure*}[ht]
\begin{center}
\centerline{\includegraphics[width=0.7\textwidth, trim={0 1.3cm 0 0.69cm},clip]{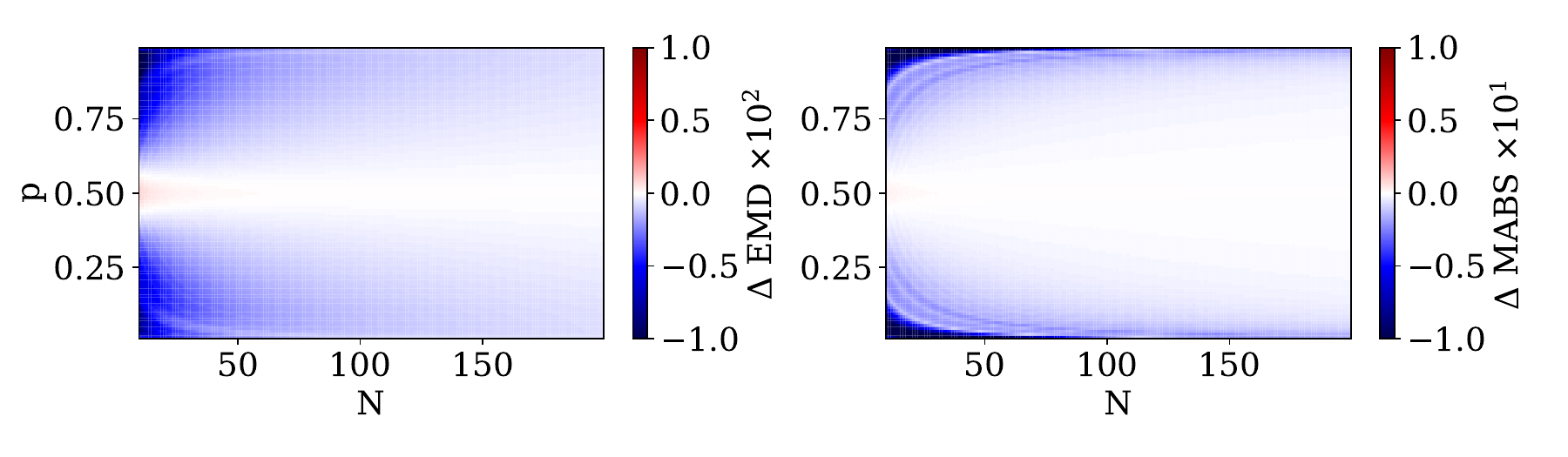}}
\centerline{\includegraphics[width=0.7\textwidth, trim={0 0.3cm 0 0.69cm},clip]{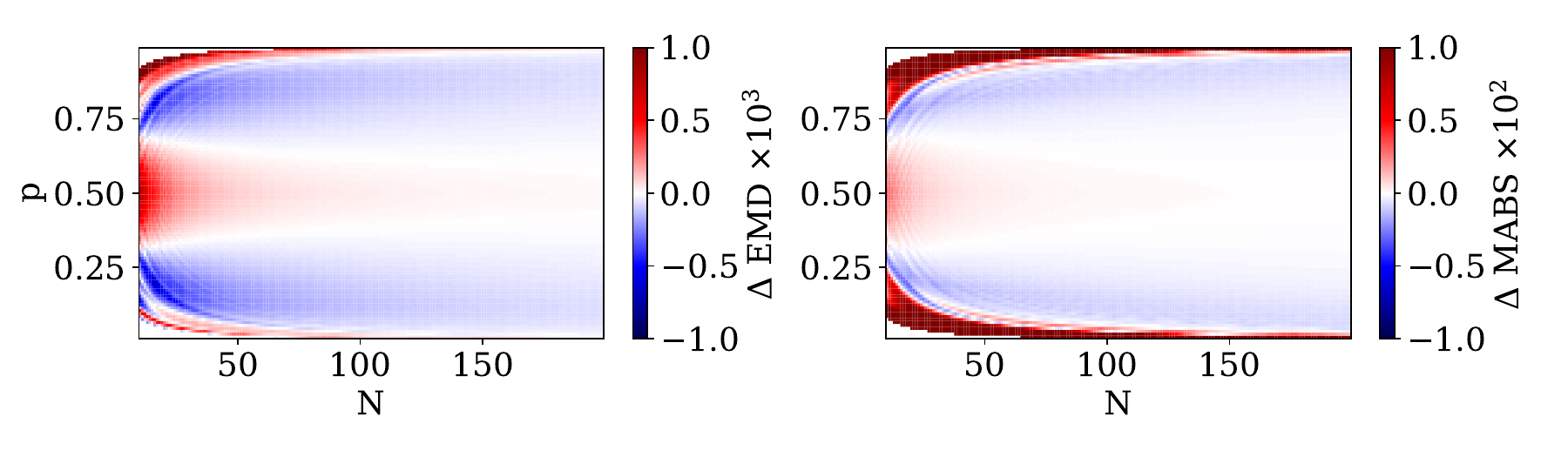}}
\caption{Comparison of the performance of the beta-normal mixture distribution introduced in~\cref{ssec:beta_norm} versus that of a normal distribution (top) and a skew-normal distribution (bottom), when approximating a binomial distribution. A comparison using the earth mover's distance is shown on the left, and one using the maximum absolute error is shown on the right. Blue and red indicate regions of the binomial distribution parameter space where the beta-normal mixture distribution outperforms and under-performs, respectively. 
The leftmost corners of the bottom plots are empty because for those regions of parameter space, we were unable to obtain numerical solutions for the skew-normal distribution.}
\label{fig:mixture_vs_norm}
\end{center}
\end{figure*}

We now compare the performance of the beta-normal mixture distribution introduced in this work, a skew-normal distribution, and a normal distribution when approximating a binomial approximation using both the maximum absolute error~\cite{chang8bd310a2-bf97-36f1-b784-d8427f10b7b9, schader:doi:10.1080/00031305.1989.10475601} and the earth mover's distance~\cite{Ramdas2015OnWT}. This is shown in~\cref{fig:mixture_vs_norm}. We can see that the normal distribution is a poorer approximation for the binomial distribution than the beta-normal mixture distribution, however, there are regions of relative over- and under-performance in the parameter space when comparing the mixture and the skew-normal distributions. The chief advantage of the mixture distribution is thus the simple closed form of the distribution parameters, enabling the implementation of a model that is fully compatible with automatic differentiation in \texttt{JAX}~\cite{jax2018github}.

\subsection{Model Implementations}\label{ssec:bayesNEST_model}

\tikzstyle{block} = [circle, draw, fill=white, 
    text width=1.2em, text centered, rounded corners, minimum size=0.1cm, font=\fontsize{8}{8}\selectfont]
\tikzstyle{obs} = [circle, draw, fill=lightgray, 
    text width=1.2em, text centered, rounded corners, minimum size=0.1cm, font=\fontsize{8}{8}\selectfont]
\tikzstyle{arrow} = [thick,->,>=stealth]
\tikzstyle{rct}=[rectangle,draw,thin,fill=white]

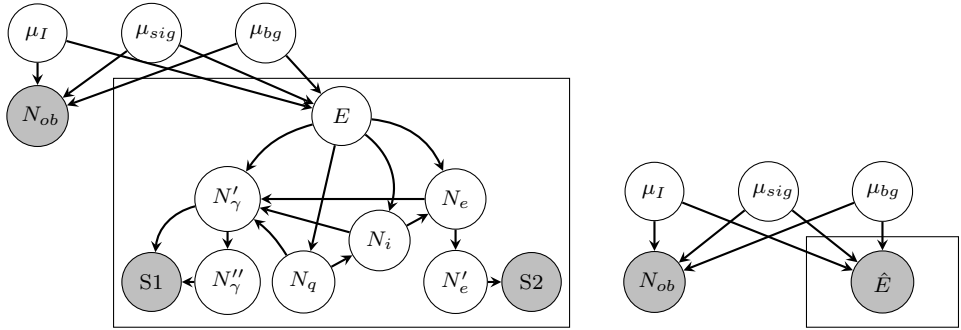
\begin{figure}[ht]
    \centering
    \begin{subfigure}[b]{0.5\textwidth}
    \begin{tikzpicture}[scale=1, node distance = 2.5cm, auto]
        \node at (0,0) [block] (mu_I) {$\mu_{I}$};
        \node at (1.5cm,0) [block] (mu_sig) {$\mu_{sig}$};
        \node at (3cm,0) [block] (mu_bg) {$\mu_{bg}$};
        \node at (0cm, -1.1*1cm) [obs] (N_obs) {$N_{ob}$};
        \node at (4cm, -1.1*1cm) [block] (E) {$E$};
        \node at (3.5cm, -1.1*3cm) [block] (N_q) {$N_q$};
        \node at (4.5cm, -1.1*2.5cm) [block] (N_i) {$N_i$};
        
        \node at (2.5cm, -1.1*2cm) [block] (Nprime_gamma) {$N'_\gamma$};
        \node at (2.5cm, -1.1*3cm) [block] (Ndprime_gamma) {$N''_\gamma$};
        \node at (1.5cm, -1.1*3cm) [obs] (S1) {S1};
        
        \node at (5.5cm, -1.1*2cm) [block] (N_e) {$N_e$};
        \node at (5.5cm, -1.1*3cm) [block] (Nprime_e) {$N'_e$};
        \node at (6.5cm, -1.1*3cm) [obs] (S2) {S2};

        \draw [draw=black] (1cm, -3.9cm) rectangle ++(6cm, 3.3cm);
        
        \draw [arrow] (mu_I) edge node []{} (N_obs);
        \draw [arrow] (mu_sig) edge node []{} (N_obs);
        \draw [arrow] (mu_bg) edge node []{} (N_obs);

        \draw [arrow] (mu_I) edge node []{} (E);
        \draw [arrow] (mu_sig) edge node []{} (E);
        \draw [arrow] (mu_bg) edge node []{} (E);

        \draw [arrow] (E) edge node []{} (N_q);
        \draw [arrow] (N_q) edge node []{} (N_i);
        \draw [arrow] (E) edge[bend left = 35] node []{} (N_i);
        \draw [arrow] (N_i) edge node []{} (Nprime_gamma);
        \draw [arrow] (N_q) edge[bend right = 10] node []{} (Nprime_gamma);
        \draw [arrow] (E) edge[bend right = 20] node []{} (Nprime_gamma);
        
        \draw [arrow] (N_i) edge node []{} (N_e);
        \draw [arrow] (E) edge[bend left = 30] node []{} (N_e);
        \draw [arrow] (N_e) edge node []{} (Nprime_gamma);
        \draw [arrow] (N_e) edge node []{} (Nprime_e);
        \draw [arrow] (Nprime_e) edge node []{} (S2);

        \draw [arrow] (Nprime_gamma) edge node []{} (Ndprime_gamma);
        \draw [arrow] (Ndprime_gamma) edge node []{} (S1);
        \draw [arrow] (Nprime_gamma) edge[bend right=35] node []{} (S1);
    \end{tikzpicture}
    \end{subfigure}
    \begin{subfigure}[b]{0.4\textwidth}
    \begin{tikzpicture}[scale=1, node distance = 2.5cm, auto]
        \node at (0,0) [block] (mu_I) {$\mu_{I}$};
        \node at (1.5cm,0) [block] (mu_sig) {$\mu_{sig}$};
        \node at (3cm,0) [block] (mu_bg) {$\mu_{bg}$};
        \node at (0cm, -1.2cm) [obs] (N_obs) {$N_{ob}$};
        \node at (3cm, -1.2cm) [obs] (E) {$\hat{E}$};

        \draw [draw=black] (2cm, -1.8cm) rectangle ++(2cm, 1.2cm);
        
        \draw [arrow] (mu_I) edge node []{} (N_obs);
        \draw [arrow] (mu_sig) edge node []{} (N_obs);
        \draw [arrow] (mu_bg) edge node []{} (N_obs);

        \draw [arrow] (mu_I) edge node []{} (E);
        \draw [arrow] (mu_sig) edge node []{} (E);
        \draw [arrow] (mu_bg) edge node []{} (E);
    \end{tikzpicture}
    \end{subfigure}
    \caption{Left: Graphical representation of bayesNEST model. A plate, indicated by the rectangle, is used to represent the parts of the model that represent the emission model described in~\cref{ssec:NEST}. Right: Graphical representation of the model for energy spectrum fit. The only variable within the plate is the estimated energy. Observed variables are shown in grey.}
    \label{fig:bayesNEST}
\end{figure}

A graphical representation of the model introduced in this work, termed bayesNEST, is shown in~\cref{fig:bayesNEST}. 
This model is implemented in \texttt{numpyro}~\cite{bingham2019pyro, phan2019composable}. The NEST emissions model (see~\cref{ssec:NEST}) is represented within the plate, as variables within the NEST emissions model are repeated for each event. $\mu_I$ represents the ~\isotope{I}{125} rate (see~\cref{ssec:DEC}) and is not used for the line search described in~\cref{ssec:line_search}. $\mu_{sig}$ is the signal, and $\mu_{bg}$ is the rate of background events. $N_{obs}$ is the total number of events observed, and is given a Poisson likelihood where $\lambda=\sum_i \mu_i$.
$E$ has a prior parameterised by the signal rate parameters. In the case of a line search, the prior corresponds to a mixture distribution with two components: a uniform distribution and a normal distribution, corresponding to the uniform background and the signal respectively. The weights of the mixture distribution are given by the ratio of $\mu_{bg}$ to $\mu_{sig}$, normalised such that the weights sum to 1. For the double-electron capture measurement, another normal distribution is added, corresponding to the \isotope{I}{125} background (see~\cref{ssec:DEC}). Mono-energetic components are represented using normal distributions with standard deviations much smaller than the energy resolution, instead of a delta function, so that the model has a smooth and continuous derivative.

We benchmark inference with the bayesNEST model (see~\cref{fig:bayesNEST}) against an energy spectrum fit. For this model, the observable is an estimator of energy, $\hat{E} = W({S1}/{g_1} + {S2}/{g_2})$,
where $W$ is the workfunction, $g_1$ is the average signal amplitude observed from a single emitted photon, and $g_2$ is the average signal amplitude observed from a single S2 electron. The prior for $\hat{E}$ is defined as a mixture distribution where the normally-distributed components have a larger standard deviation, corresponding to the standard deviation of the estimator $\hat{E}$ given a mono-energetic component. This energy resolution is estimated for each mono-energetic component by generating $5\times10^4$ events using the NEST simulator, computing $\hat{E}$ for each event, and then computing the corrected sample standard deviation.

\section{Results}
\subsection{Mono-energetic Signal Search with Known Background Shape}\label{ssec:line_search}

\begin{figure}[ht]
\centering
\begin{subfigure}[b]{0.38\textwidth}
\includegraphics[width=\textwidth, trim={0 0 0 0.1cm},clip]{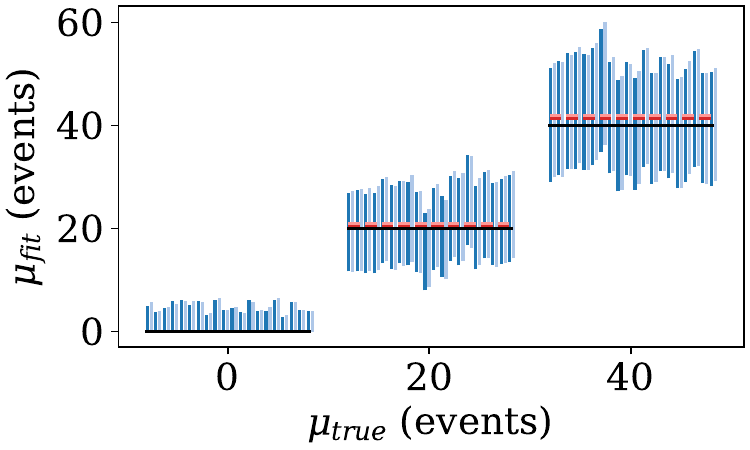}
\end{subfigure}
\begin{subfigure}[b]{0.38\textwidth}
\includegraphics[width=\textwidth, trim={0 0 0 0.1cm},clip]{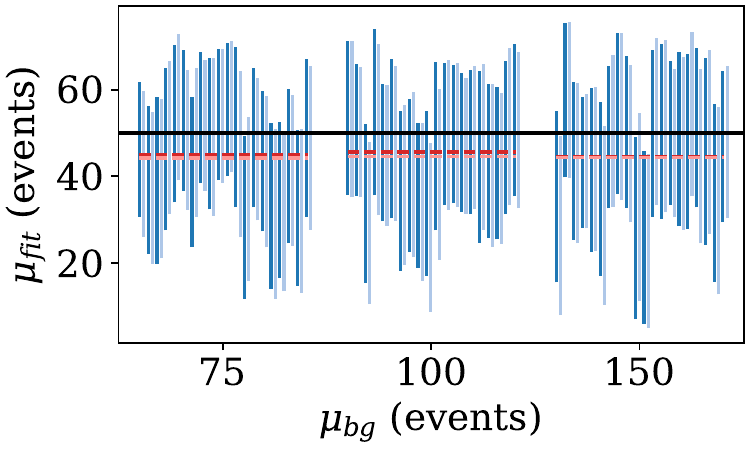}
\end{subfigure}
\caption{Left: $90\%$ credible intervals for the signal rate parameter for the mono-energetic signal search. 20 intervals are computed for each of $\mu_{true} \in \{0, 20, 40\}$. Right: $90\%$ credible intervals for the signal rate parameter for the double-electron capture search. 20 intervals are computed for each of $\mu_{bg} \in \{75, 100, 150\}$. Intervals computed using bayesNEST are shown in dark blue, and those computed from a spectral fit are shown in light blue. The mean from all samples for each set of 20 runs is shown in red and pink for bayesNEST and for the spectral fit, respectively.}
\label{fig:line_fit}
\end{figure}

We first compare the difference between inference using a simple spectral fitting model and the bayesNEST model (see~\cref{ssec:bayesNEST_model}) using a search for a mono-energetic signal with a flat background. This simulates searches for dark matter or physics beyond the standard model using the electronic recoil channel in liquid xenon detectors, such as those conducted in~\cite{XENON:2020rca} and~\cite{XENON:2022ltv}.

\begin{table}[ht]
    \centering
    \caption{Left: The mean $90\%$ credible intervals from bayesNEST $\left(\mathbb{E}[\sigma_{90\%}]\right)$, spectral fitting $\left(\mathbb{E}[\sigma'_{90\%}]\right)$, and the mean relative difference in interval size $\left(\mathbb{E}\left[1-{\sigma'_{90\%}}/{\sigma_{90\%}}\right]\right)$ for the mono-energetic signal search. Right: The mean $90\%$ credible intervals from bayesNEST $\left(\mathbb{E}[\sigma_{90\%}]\right)$, spectral fitting $\left(\mathbb{E}[\sigma'_{90\%}]\right)$, and the mean relative difference in interval size $\left(\mathbb{E}\left[1-{\sigma'_{90\%}}/{\sigma_{90\%}}\right]\right)$ for the double-electron capture search.}\label{tab:interval_size}
    \begin{tabular}{c c c c}
        \hline
        \rule{0pt}{1.5em}
        $\mu_{true}$ \qquad & $\mathbb{E}[\sigma_{90\%}]$ \qquad& $\mathbb{E}[\sigma'_{90\%}]$ \qquad& $\mathbb{E}\left[1-\frac{\sigma'_{90\%}}{\sigma_{90\%}}\right]$\\[5pt]
        \hline
        \hline
        0 & $4.8 \pm 0.2$ & $4.9 \pm 0.2$ & $3.6 \pm 0.9 \%$\\
        20 & $16.1 \pm 0.1$ & $16.4 \pm 0.1$ & $1.6 \pm 0.4 \%$\\
        40 & $22.2 \pm 0.1$ & $22.3 \pm 0.1$ & $0.5 \pm 0.3 \%$\\
        \hline
    \end{tabular}
    \begin{tabular}{c c c c}
        \hline
        \rule{0pt}{1.5em}
        $\mu_{bg}$ \qquad & $\mathbb{E}[\sigma_{90\%}]$ \qquad& $\mathbb{E}[\sigma'_{90\%}]$ \qquad& $\mathbb{E}\left[1-\frac{\sigma'_{90\%}}{\sigma_{90\%}}\right]$\\[5pt]
        \hline
        \hline
        75 & $34.6 \pm 0.6$ & $35.0 \pm 0.6$ & $0.9 \pm 0.7 \%$\\
        100 & $35.2 \pm 0.6$ & $35.8 \pm 0.6$ & $2.0 \pm 0.7 \%$\\
        150 & $37.8 \pm 0.7$ & $38.5 \pm 0.7$ & $1.9 \pm 0.8 \%$\\
        \hline
    \end{tabular}
\end{table}

We simulated background data with a flat spectrum between $50\unit{keV}$ and $350\unit{keV}$ and a rate of 50 events over the interval using \texttt{NEST} version 2.3.11~\cite{szydagis_2023_7577399} and \texttt{nestpy}~\cite{farrell_2023_7552304}. A search for a mono-energetic peak at an energy of $200\unit{keV}$ is conducted on events selected to have estimated energies of between $100\unit{keV}$ and $300\unit{keV}$.
The NUTS sampler implemented in \texttt{numpyro}~\cite{phan2019composable, bingham2019pyro} is used. This is repeated 20 times each for true signal rates ($\mu_{true}$) of $0$, $20$, and $40$ events. These experiments were run using 8 CPU cores, and took less than 1 day. The results are shown on the left in~\cref{fig:line_fit} and~\cref{tab:interval_size}.
We can see that inference using bayesNEST has reduced variance, and that the mean is closer to the true value $\mu_\mathrm{true}$.

\subsection{Searching for Rare Nuclear Decays}\label{ssec:DEC}

We then compare the difference between the two methods of inference by simulating a search for the radioactive decay of \isotope{Xe}{124} via double-electron capture. This is a challenging problem, as in addition to a smooth background component, there exists an \isotope{I}{125} background which is also mono-energetic at $67.3\unit{keV}$, very close to the signal we are interested in, at $64.3\unit{keV}$~\cite{XENON:2019dti}. 


For this test, we fix the signal rate to be $50$ events and the \isotope{I}{125} background rate to be $5$ events when simulating data. We consider 3 scenarios with differing rates of the flat background ($\mu_{bg}$): $75$ events, $100$ events, and $150$ events. This is defined on an energy range of $40\unit{keV}$ to $110\unit{keV}$, and for inference, data with estimated energies of between $50\unit{keV}$ and $100\unit{keV}$ are selected. These experiments were run using 8 CPU cores and took less than 1 day to complete as well. The results are shown on the right in~\cref{fig:line_fit} and~\cref{tab:interval_size}.
As with the results in~\cref{ssec:line_search}, we can see that inference with bayesNEST produces results with lower uncertainty, and the mean over 20 runs is closer to the true signal rate of $50$ events.

\section{Discussion}
Inference in liquid xenon dark matter experiments has largely been based on the use of estimators of event energy~\cite{XENON:2019dti, XENON:2020rca, XENON:2022ltv}, or comparisons of data with empirical distributions from simulated data~\cite{XENON:2018voc,XENON:2019izt}. Prior work on computing explicit event-by-event likelihoods~\cite{James:2022sgg} has relied on marginalising a discrete likelihood over a large array, requiring computationally and memory intensive enumeration of discrete latent variables. In this work, we demonstrate the use of continuous approximations to construct such a likelihood, alleviating the need for the marginalisation procedure, and enabling inference using gradient-based inference methods, such as variational inference~\cite{Hoffman2012StochasticVI, wingate2013automated} and the No-U-Turn Sampler (NUTS)~\cite{Hoffman2011TheNS}. 

The approximate procedure introduced in this work produces improvements over inference using an energy estimator, offering percent-level improvements in the uncertainty of inferred parameters. In the background-dominated regime, sensitivity is asymptotically proportional to $\mathrm{signal}/\sqrt{\mathrm{background}}$ \cite{Cowan:2010js}; as such, if both signals and background event rates are proportional to the detector target mass, the sensitivity scales as the square-root of exposure. A $4\%$ improvement such as that demonstrated in~\cref{ssec:line_search} would thus correspond to a $7\%$ increase in exposure.
The high cost of xenon~\cite{LZ:2015kxe} means that even small improvements from analysis or inference technique can represent very cost-effect investments.

In addition to improvements in inference, the use of explicit likelihoods allow for inference to easily incorporate additional information if available. For example, $N'_{\gamma}$ can be directly estimated by counting pulses in photosensors independently of the S1 amplitude~\cite{LUX:2015amk, LUX:2016ggv}. This can be easily incorporated into the model introduced in this work, either by directly making $N'_{\gamma}$ in the model an observable, or by introducing a new observable depending on $N'_{\gamma}$ while modelling for detection efficiency. Discrete measurements from pulse counting are often better at lower energies and continuous measurements based on integration of waveforms are often better at higher energies. The ability to integrate discrete and continuous measurements is also invaluable, as it would allow for a single unified scheme that can make use of both sources of information, without having a sharp transition point between the two methods that might complicate inference or data analysis.


\section*{Acknowledgements}
This work was supported by the National Science Foundation under Grant No. 2046549.
\bibliographystyle{vancouver}
\bibliography{main}

\begin{thebibliography}{10}

\bibitem{Cranmer_2020}
Cranmer K, Brehmer J, Louppe G.
\newblock The frontier of simulation-based inference.
\newblock Proceedings of the National Academy of Sciences. 2020 May;117(48):30055–30062.
\newblock Available from: \url{http://dx.doi.org/10.1073/pnas.1912789117}.

\bibitem{Baydin2018EfficientPI}
Baydin AG, Heinrich L, Bhimji W, Munk A, Gram-Hansen B, Louppe G, et~al.
\newblock Efficient Probabilistic Inference in the Quest for Physics Beyond the Standard Model.
\newblock In: Neural Information Processing Systems; 2018. p. 5459  5472.
\newblock Available from: \url{https://api.semanticscholar.org/CorpusID:49903021}.

\bibitem{Baydin_2019}
Baydin AG, Shao L, Bhimji W, Heinrich L, Meadows L, Liu J, et~al.
\newblock Etalumis: bringing probabilistic programming to scientific simulators at scale.
\newblock In: Proceedings of the International Conference for High Performance Computing, Networking, Storage and Analysis. SC ’19. ACM; 2019. p. 1  24.
\newblock Available from: \url{http://dx.doi.org/10.1145/3295500.3356180}.

\bibitem{jax2018github}
Bradbury J, Frostig R, Hawkins P, Johnson MJ, Leary C, Maclaurin D, et~al.. {JAX}: composable transformations of {P}ython+{N}um{P}y programs; 2018.
\newblock Available from: \url{http://github.com/google/jax}.

\bibitem{Hoffman2012StochasticVI}
Hoffman MD, Blei DM, Wang C, Paisley JW.
\newblock Stochastic variational inference.
\newblock J Mach Learn Res. 2012;14:1303-47.
\newblock Available from: \url{https://api.semanticscholar.org/CorpusID:5652538}.

\bibitem{wingate2013automated}
Wingate D, Weber T. Automated Variational Inference in Probabilistic Programming; 2013.

\bibitem{Hoffman2011TheNS}
Hoffman MD, Gelman A.
\newblock The No-U-turn sampler: adaptively setting path lengths in Hamiltonian Monte Carlo.
\newblock J Mach Learn Res. 2011;15:1593-623.
\newblock Available from: \url{https://api.semanticscholar.org/CorpusID:12948548}.

\bibitem{bingham2019pyro}
Bingham E, Chen JP, Jankowiak M, Obermeyer F, Pradhan N, Karaletsos T, et~al.
\newblock Pyro: Deep Universal Probabilistic Programming.
\newblock J Mach Learn Res. 2019;20:28:1-28:6.
\newblock Available from: \url{http://jmlr.org/papers/v20/18-403.html}.

\bibitem{phan2019composable}
Phan D, Pradhan N, Jankowiak M.
\newblock Composable Effects for Flexible and Accelerated Probabilistic Programming in NumPyro.
\newblock arXiv preprint arXiv:191211554. 2019.

\bibitem{XENON:2020kmp}
Aprile E, et~al.
\newblock {Projected WIMP sensitivity of the XENONnT dark matter experiment}.
\newblock JCAP. 2020;11:031.

\bibitem{LZ:2019sgr}
Akerib DS, et~al.
\newblock {The LUX-ZEPLIN (LZ) Experiment}.
\newblock Nucl Instrum Meth A. 2020;953:163047.

\bibitem{ParticleDataGroup:2022pth}
Workman RL, et~al.
\newblock {Review of Particle Physics}.
\newblock PTEP. 2022;2022:083C01.

\bibitem{Aalbers:2022dzr}
Aalbers J, et~al.
\newblock {A next-generation liquid xenon observatory for dark matter and neutrino physics}.
\newblock J Phys G. 2023;50(1):013001.

\bibitem{XENON:2019dti}
Aprile E, et~al.
\newblock {Observation of two-neutrino double electron capture in $^{124}$Xe with XENON1T}.
\newblock Nature. 2019;568(7753):532-5.

\bibitem{XENON:2022evz}
Aprile E, et~al.
\newblock {Double-Weak Decays of $^{124}$Xe and $^{136}$Xe in the XENON1T and XENONnT Experiments}.
\newblock Phys Rev C. 2022;106(2):024328.

\bibitem{XENON:2020rca}
Aprile E, et~al.
\newblock {Excess electronic recoil events in XENON1T}.
\newblock Phys Rev D. 2020;102(7):072004.

\bibitem{XENON:2022ltv}
Aprile E, et~al.
\newblock {Search for New Physics in Electronic Recoil Data from XENONnT}.
\newblock Phys Rev Lett. 2022;129(16):161805.

\bibitem{XENON:2019izt}
Aprile E, et~al.
\newblock {XENON1T dark matter data analysis: Signal and background models and statistical inference}.
\newblock Phys Rev D. 2019;99(11):112009.

\bibitem{James:2022sgg}
James RS, Palmer J, Kaboth A, Ghag C, Aalbers J.
\newblock {FlameNEST: explicit profile likelihoods with the Noble Element Simulation Technique}.
\newblock JINST. 2022;17(08):P08012.

\bibitem{LZ:2015kxe}
Akerib DS, et~al.
\newblock {LUX-ZEPLIN (LZ) Conceptual Design Report}. 2015 9.

\bibitem{LZ:2022lsv}
Aalbers J, et~al.
\newblock {First Dark Matter Search Results from the LUX-ZEPLIN (LZ) Experiment}.
\newblock Phys Rev Lett. 2023;131(4):041002.

\bibitem{szydagis_2023_7577399}
Szydagis M, Brown E, Carrara N, Kamaha AC, Kozlova ES, McKinsey DN, et~al.. Noble Element Simulation Technique. Zenodo; 2023.
\newblock Available from: \url{https://doi.org/10.5281/zenodo.7577399}.

\bibitem{Faham:2015kqa}
Faham CH, Gehman VM, Currie A, Dobi A, Sorensen P, Gaitskell RJ.
\newblock {Measurements of wavelength-dependent double photoelectron emission from single photons in VUV-sensitive photomultiplier tubes}.
\newblock JINST. 2015;10(09):P09010.

\bibitem{Gebhardt:127bc0fe-1b1f-35af-bccd-7687f9f56c9e}
Gebhardt F.
\newblock Some Numerical Comparisons of Several Approximations to the Binomial Distribution.
\newblock Journal of the American Statistical Association. 1969;64(328):1638-46.
\newblock Available from: \url{http://www.jstor.org/stable/2286095}.

\bibitem{chang8bd310a2-bf97-36f1-b784-d8427f10b7b9}
Chang CH, Lin JJ, Pal N, Chiang MC.
\newblock A Note on Improved Approximation of the Binomial Distribution by the Skew-Normal Distribution.
\newblock The American Statistician. 2008;62(2):167-70.
\newblock Available from: \url{http://www.jstor.org/stable/27643999}.

\bibitem{Walck:1996cca}
Walck C.
\newblock {Hand-book on statistical distributions for experimentalists}; 1996.

\bibitem{hoffalma991033186282805251}
Hoff PD.
\newblock A first course in Bayesian statistical methods Peter D. Hoff.
\newblock 1st ed. Springer texts in statistics. New York, N.Y: Springer; 2009.

\bibitem{schader:doi:10.1080/00031305.1989.10475601}
Schader M, Schmid F.
\newblock Two Rules of Thumb for the Approximation of the Binomial Distribution by the Normal Distribution.
\newblock The American Statistician. 1989;43(1):23-4.
\newblock Available from: \url{https://www.tandfonline.com/doi/abs/10.1080/00031305.1989.10475601}.

\bibitem{Ramdas2015OnWT}
Ramdas A, Trillos NG, Cuturi M.
\newblock On Wasserstein Two-Sample Testing and Related Families of Nonparametric Tests.
\newblock Entropy. 2015;19:47.
\newblock Available from: \url{https://api.semanticscholar.org/CorpusID:7725237}.

\bibitem{farrell_2023_7552304}
Farrell S, Tunnell C, Angevaare JR, Rischbieter GRC, Carrara N, mszydagis, et~al.. {NESTCollaboration/nestpy: Sync with NESTv2.3.12beta}. Zenodo; 2023.
\newblock Available from: \url{https://doi.org/10.5281/zenodo.7552304}.

\bibitem{XENON:2018voc}
Aprile E, et~al.
\newblock {Dark Matter Search Results from a One Ton-Year Exposure of XENON1T}.
\newblock Phys Rev Lett. 2018;121(11):111302.

\bibitem{Cowan:2010js}
Cowan G, Cranmer K, Gross E, Vitells O.
\newblock {Asymptotic formulae for likelihood-based tests of new physics}.
\newblock Eur Phys J C. 2011;71:1554.
\newblock [Erratum: Eur.Phys.J.C 73, 2501 (2013)].

\bibitem{LUX:2015amk}
Akerib DS, et~al.
\newblock {Tritium calibration of the LUX dark matter experiment}.
\newblock Phys Rev D. 2016;93(7):072009.

\bibitem{LUX:2016ggv}
Akerib DS, et~al.
\newblock {Results from a search for dark matter in the complete LUX exposure}.
\newblock Phys Rev Lett. 2017;118(2):021303.

\end{thebibliography}

\end{document}